\documentclass[aps,pre,groupedaddress,10pt]{revtex4-1}

\renewcommand{\d}{\mathrm{d}}
\usepackage{graphicx,bm,amsmath,color}

\graphicspath{{./figures/}}

\begin{document}


\title{Soft beams: when capillarity induces axial compression}


\author{S. Neukirch$^{1,2}$}
\author{A. Antkowiak$^{1,2}$}
\author{J.-J. Marigo$^3$}
\affiliation{
$^1$CNRS, UMR 7190, Institut Jean Le Rond d'Alembert, F-75005 Paris, France.\\
$^2$UPMC Univ Paris 06, UMR 7190, Institut Jean Le Rond d'Alembert, F-75005 Paris, France.\\
$^3$CNRS, Ecole Polytechnique, UMR 7649, Lab. M\'eca. Solides, F-91128 Palaiseau Cedex, France
}


\date{\today}

\begin{abstract}
We study the interaction of an elastic beam with a liquid drop in the case where bending and extensional effects are both present. We use a variational approach to derive equilibrium equations and constitutive relation for the beam.
This relation is shown to include a term due to surface energy in addition of the classical Young's modulus term, leading to a modification of Hooke's law.
 At the triple point where solid, liquid, and vapor phases meet we find that the external force applied on the beam is parallel to the liquid-vapor interface. Moreover, in the case where solid-vapor and solid-liquid interface energies do not depend on the extension state of the beam, we show that the extension in the beam is continuous at the triple point and that the wetting angle satisfy the classical Young-Dupr\'e relation.
\end{abstract}


\maketitle

\section{Introduction}
%
%
%
%
As for other surface effects, capillarity typically comes into play at small scales. Wetting phenomena and capillary effects have been classically studied in the context of instabilities and morphogenesis with menisci, bubbles, drops, foams in the leading roles \cite{degennes+al:2003}.
More recently interactions between these structures and elastic solids have been considered and when elastic forces are of the same order of magnitude as surface tension rich behaviors have been found \cite{Roman-Bico:Elasto-capillarity:-deforming-an-elastic:2010}.
As surface tension is a rather small force it has to be applied to flexible or soft systems, typically slender structures with low bending rigidity. In such setups, aggregation \cite{bico+al:2004,Py-Bastien-3D-aggregation-of-wet-fibers-2007}, folding \cite{Py-Reverdy-Capillary-Origami:-Spontaneous-2007}, and snapping \cite{Fargette-Antkowiak:Elastocapillary-snapping:2013} have been demonstrated.
In the case of low Young's modulus, surface tension may as well induce compression and wrinkling \cite{Huang-Juszkiewicz:Capillary-Wrinkling-of-Floating:2007,Mora2010Capillarity-Dri}.
Apart from biological systems where capillarity and/or elasticity might be key players \cite{cohen+mahadevan:2003}, elastocapillary interactions lie at the core of several engineering applications for example in the field of micro and nano-systems \cite{Syms-Yeatman:Surface-tension-powered-self-assembly:2003,De-Volder-Hart:Engineering-Hierarchical-Nanostructures:2013}.
In order to solve for the deformation of an elastic solid one has to know the external applied forces and boundary conditions. In the case of a liquid drop sitting on a elastic plate or strip, pressure and meniscus forces act in concert and induce bending and compression \cite{Antkowiak-Audoly:Instant-fabrication-and-selection:2011}.
The intensity and direction of such forces, as well as the resulting compression, have recently been discussed \cite{Marchand-Das:Capillary-Pressure-and-Contact:2012}.

Here we derive, from energy principles, the equations that rule the equilibrium of such drop-strip systems, and we highlight unusual (constitutive) relations between forces and deformations in the presence of capillarity.
We start in Section \ref{section:ext-seule} with purely extensional setups. We first consider the case of a beam subject to both surface tension and external end load, and we show how the classical constitutive relation is modified. We then study the case of a beam in interaction with two symmetric droplets and compute the applied force at menisci together with the resulting extension in the beam.
In Section \ref{section:flexion-ext} we consider the case of a single drop deposited on the beam, a setup where both extension and bending occur. We compute the force at the meniscus and the resulting extension in the beam.
In the Appendix we consider the case where solid-liquid and solid-vapor interface energies depend on the strain state in the beam, and we list the differences and similarities with results of Section \ref{section:ext-seule}.

%
%
%
%
%
%
\section{Beams in pure extension} \label{section:ext-seule}

\subsection{Compression due to the environment} \label{section:comp-enviro}
%
%
%
%

\begin{figure}[ht]
    \centering
    \includegraphics{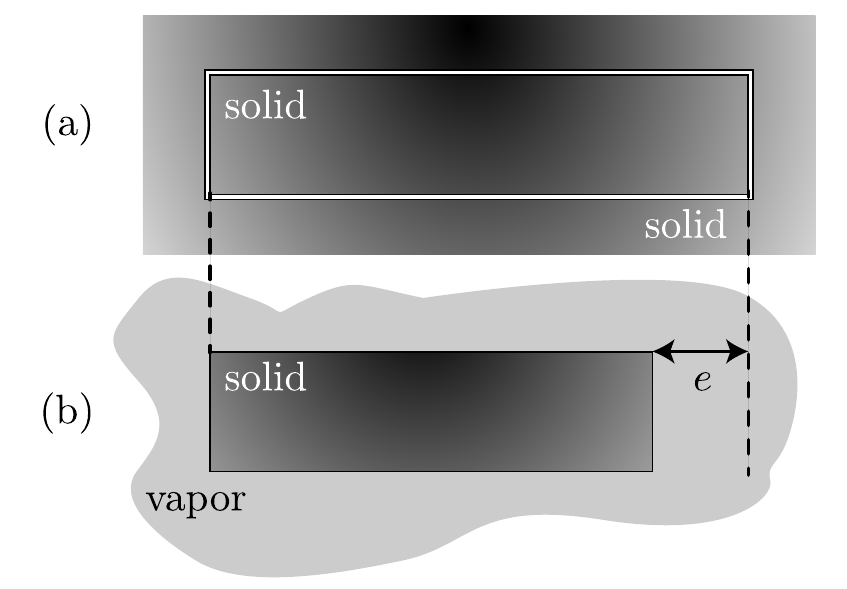}
    \caption{(a) A beam surrounded by its own material. In this abstract configuration there is no interface energy. (b) A beam (solid phase S) surrounded by air (vapor phase V). In this configuration the interface energy is $\gamma_{\text{sv}}$ per unit area.}
    \label{fig1}
\end{figure}

We start with considering a beam made out of a solid material of low Young's modulus $E$, typically $E \sim $ kPa \citep{Mora2010Capillarity-Dri}. For such low Young's modulus the stretching energy
\begin{equation}
V_\text{e} = \frac{1}{2} \int_0^L EA \, e^2(s)  \d s
\label{equa:Velas}
\end{equation}
is of the same order of magnitude that the surface energy
\begin{equation}
V_\text{s} = \gamma_{\text{sv}} P \int_0^L [1+e(s)] \d s
\label{equa:Vsurface}
\end{equation}
where $L$ is the length of the beam in the reference state, $A$ (respectively $P$) its cross-section area (resp. perimeter), $e(s)$ the extension strain with $e>0$ (resp. $e<0$) corresponding to extension (resp. compression), $s$ the arc-length along the beam in the reference state, and $\gamma_{\text{sv}}$ the surface energy corresponding to the solid-vapor interface \footnote{ In Sections  \ref{section:ext-seule} and \ref{section:flexion-ext} surface energies $\gamma$ will be assumed to be independent of the extension strain $e$. The case where $\gamma=\gamma(e)$ is treated in Appendix \ref{appendiceA}}.
In a configuration where such a solid beam lies at equilibrium in a vapor phase, the extension $e$ is uniform and is obtained by minimizing the total energy $V(e)=V_\text{e}+V_\text{s}$. Imposing $V'(e)=0$ yields
\begin{equation}
e = - \frac{\gamma_{\text{sv}} P}{EA} <0
\end{equation}
that is $e=-0.02$ (2\% compression) for a beam with $E=10$ kPa, $\gamma_{\text{sv}}=0.01$ N/m, and circular cross-section of radius $h=0.1$ mm. Such a beam is compressed due to surface tension, that is its current length $(1+e)L$ is $2\%$ less that in a situation where the beam would be surrounded by a solid phase of its own material, as in Fig.~\ref{fig1}-(a).

\subsection{Tension-extension constitutive relation} \label{section:const-rel}
%
%
%
%
\begin{figure}[ht]
    \centering
    \includegraphics{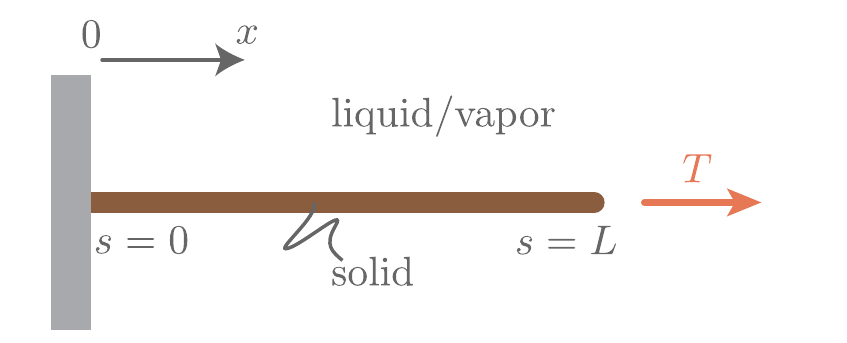}
    \caption{A beam (solid phase S) surrounded by a phase $i$ (liquid or vapor). The beam is subjected to an external tension $T$.}
    \label{fig2}
\end{figure}
We now consider a beam (solid phase S) surrounded by a phase $i$ (in the following this phase will either be liquid $i=\ell$ or vapor $i=V$). The beam is anchored at one extremity and subjected to an external tension at the other end, see Fig.~\ref{fig2}. As in the former section the beam  has Young's modulus $E$, cross-section area $A$ and perimeter $P$, reference length $L$. To the internal energy $V_\text{e}+V_\text{s}$ (see Eqs. (\ref{equa:Velas}) and (\ref{equa:Vsurface})) we add the work done by the external load $T$: $W_T= -T(x(L)-x(0))=- T \int_0^L x'(s) \d s$ to obtain the total potential energy $V=V_\text{e}+V_\text{s}+W_T$ that depends on two unknown functions $e(s)$ and $x(s)$. Definition of strain $e(s)=x'(s)-1$ implies that these two unknowns are linked by a continuous constraint. We therefore introduce a continuous Lagrange multiplier $\nu(s)$ and work with
\begin{equation}
{\cal L}(x(s),e(s))=V_\text{e}+V_\text{s}+W_T + \int_0^L \nu(s) \, [x'(s) - (1+e(s)) ] \d s
\end{equation}
We introduce perturbations $x \to x + \varepsilon \bar{x}$, $e \to e + \varepsilon \bar{e}$
and we compute the expansion
\begin{equation}
{\cal L}(x + \varepsilon \bar{x}, e + \varepsilon \bar{e}) = {\cal L}(x,e) + \varepsilon \left. \frac{\d {\cal L}}{\d \varepsilon} \right|_{\varepsilon=0}+ \ldots
\end{equation}
A necessary condition for $V$ to be minimum at $(x,e)$ under the constraint $x'(s)=1+e(s)$ is that the first variation $(\d {\cal L} / \d \varepsilon)_{\varepsilon=0}$ vanishes at $(x,e)$ for all $(\bar{x},\bar{e})$. After derivation and integration by parts, we obtain
\begin{equation}
\left. \frac{\d {\cal L}}{\d \varepsilon} \right|_{\varepsilon=0} = \int_0^L (EA e+P \gamma_{\text{s}i}-\nu)\,  \bar{e}(s)\d s - \int_0^L \nu'(s) \, \bar{x}(s) \d s - \bigl[ (T-\nu) \bar{x} \bigr]_0^L
\label{equa:1st-varia-simple-case}
\end{equation}
with $\gamma_{\text{s}i}$ the surface energy for the interface between the solid beam and the surrounding phase $i$. Boundary condition $x(0)=0$ yields $\bar{x}(0)=0$ but $\bar{x}(L)$ is arbitrary. Consequently requiring (\ref{equa:1st-varia-simple-case}) to vanish for all $(\bar{x}(s),\bar{e}(s))$ brings equations:
\begin{eqnarray}
\nu'(s)&=& 0 \label{equa:nu} \\
\nu(s)&=& EA \, e(s) + P \gamma_{\text{s}i}\label{equa:rel-nu-e}
\end{eqnarray}
and the natural boundary condition:
\begin{equation}
\nu(L)=T \label{EQ6}
\end{equation}
In the light of this last relation, we interpret the Lagrange multiplier $\nu(s)$ as the
beam internal tension $N(s)$.
Using (\ref{equa:rel-nu-e}), we obtain the following constitutive relation 
\begin{equation}
N(s) = EA \,  e(s) + P \gamma_{\text{s}i} \label{EQ7}
\end{equation}
between the extension $e(s)$ and internal tension $N(s)$.
The total tension $N$ is seen as the sum of the bulk force $EA e(s)$ and the surface stress $P \gamma_{\text{s}i}$.
This is comparable to Hooke's law in thermoelasticity where local stress is created by both strain and temperature change. The deformation due to interface energy is analogous to the classic deformation observed when heating a beam away from its fabrication temperature, the surface energy $\gamma$ playing the role of a negative thermal expansion coefficient \citep{landau}.

Note that the situation of Section \ref{section:comp-enviro} is regained by setting $T=0$.

\subsection{Force jump at the contact line} \label{section:force-jump}
%
%
%
%
\begin{figure}[ht]
    \centering
    \includegraphics{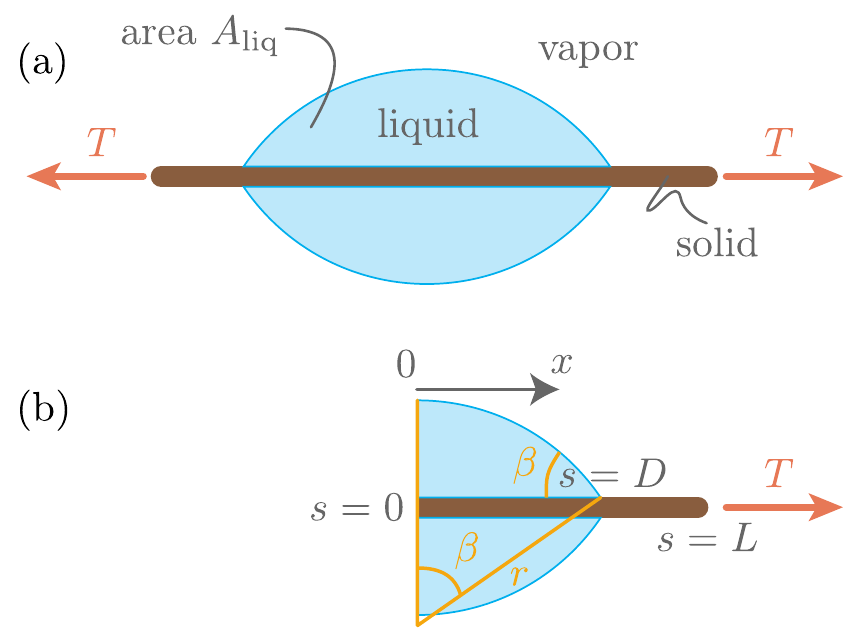}
    \caption{(a) An extensible beam bears two liquid drops. In addition the beam is subjected to an external tension $T$. (b) Using symmetry we study the right half problem and fix the midpoint of the beam $s=0$ at the origin.}
    \label{fig3}
\end{figure}
Next we consider a beam subjected to capillary interactions. A beam with rectangular cross-section (thickness $h$, width $w \gg h$) bears a liquid drop on its upper surface and a similar drop on its lower surface, see figure \ref{fig3}-(a). An external tension is applied at both extremities. We look for the equilibrium equations for such a beam under the following assumptions.
Due to the large aspect ratio $w \gg h$ of the cross-section we work in a two-dimensional setup, considering the problem invariant along the third dimension. Consequently we neglect end effects, write $P=2 w$, and consider the liquid-vapor interfaces to be cylinder arcs, see \cite{Rivetti-Neukirch:Instabilities-in-a-drop-strip-system::2012}. 
We also assume that the simultaneous presence of two liquid drops prevents the beam from bending and we only deal with extensional deformations. We break the translation invariance by fixing the beam middle point $s=0$ at the origin: $x(s=0)=0$. Consequently we solve the planar half problem of figure \ref{fig3}-(b).

As capillary forces will be applied on the beam at the contact line, where the three phases meet, we anticipate the possibility of a discontinuity in the extension and therefore introduce the extension $e_\text{i}(s)$ inside the drops, and $e_\text{o}(s)$ outside the drops. The stretching energy of the beam is now:
\begin{equation}
V_\text{e}= \frac{1}{2} \int_0^D EA e_\text{i}^2(s) \d s + \frac{1}{2} \int_D^L EA e_\text{o}^2(s) \d s \label{equa-stretch-energy}
\end{equation}
The work of the external load is $W_T=- T x(L)$, with $x(L)=\int_0^D x_\text{i}'(s) \d s+\int_D^L x_\text{o}'(s) \d s$.
The sum of the interface energies is:
\begin{eqnarray}
V_\text{s} & = & 2 w \gamma_{\text{s} \ell} \, x(D) + 2 w \gamma_{\text{sv}} [x(L)-x(D)] + 2 w \gamma_{\ell \text{v}} r \beta \nonumber  \\
&=& 2 w \gamma_{\text{s} \ell} \int_0^D  [1+e_\text{i}(s)] \d s + 2 w \gamma_{\text{sv}} \int_D^L [1+e_\text{o}(s)] \d s + 2 w \gamma_{\ell \text{v}} r \beta
\end{eqnarray}
where $r$ is the radius of both circular liquid-vapor interfaces, and $\beta$ is the wetting angle.

We want to minimize the total potential energy $V=V_\text{e}+V_\text{s}+W_T$ under the following constraints. First, as in the previous section we have relations linking unknown functions $x(s),e_\text{i}(s),e_\text{o}(s)$: for $s$ in $(0,D)$, $x_\text{i}'(s)=1+ e_\text{i}(s)$, and for $s$ in $(D,L)$, $x_\text{o}'(s)=1+e_\text{o}(s)$. Second, the total liquid volume ${\cal V}= A_{\mathrm{liq}} w$ being fixed, we have:
\begin{equation}
w r^2 (\beta - \sin \beta \cos \beta) = A_{\mathrm{liq}} w
\end{equation}
where $A_{\mathrm{liq}}$ is the area of the surface lying in between the beam and the liquid-vapor interface. Finally the position of the contact line imposes:
\begin{equation}
x(D) = r \sin \beta \label{equa-xd-contraint}
\end{equation}
We therefore will compute the first variation of:
\begin{eqnarray}
{\cal L} &=& V_\text{e}+V_\text{s}+W_T - \mu w \left[  r^2 (\beta - \sin \beta \cos \beta) - A_{\mathrm{liq}} \right] -\eta \left[ \int_0^D x'_\text{i}(s) \d s  - r \sin \beta \right] + \nonumber \\
&&  \int_0^D \nu_\text{i} [x'_\text{i}-(1+e_\text{i})] \d s + \int_D^L \nu_\text{o} [x'_\text{o}-(1+e_\text{o})] \d s
\end{eqnarray}
where $(D, \beta, r,e_\text{i}(s),e_\text{o}(s),x_\text{i}(s),x_\text{o}(s))=X$ are unknowns and $(\mu, \eta, \nu_\text{i}(s),\nu_\text{o}(s))$ Lagrange multipliers.
We therefore introduce the perturbation $X \to X+\varepsilon \bar{X}$, and we look for the conditions for which $\left. \d {\cal L}(X+\varepsilon \bar{X}) / \d \varepsilon \right|_{\varepsilon=0}=0$. We perform integration by parts to get rid of $\bar{x}_\text{i}'(s)$ and $\bar{x}_\text{o}'(s)$ terms, and we obtain
\begin{eqnarray}
\left. \frac{\d {\cal L}}{\d \varepsilon} \right|_{\varepsilon=0}
&=& \bar{\beta} \left[ 2 w \gamma_{\ell \text{v}} r - \mu w r^2 (1-\cos 2 \beta) + \eta r \cos \beta \right]  \nonumber \\
&& + \bar{r} \left[ 2 w \gamma_{\ell \text{v}} \beta - \mu w r ( 2 \beta-\sin 2 \beta) + \eta \sin \beta \right]  \nonumber \\
&& + \int_0^D \left(EA e_\text{i} + 2w \gamma_{\text{s} \ell} - \nu_\text{i} \right) \bar{e}_\text{i}\d s 
+\int_D^L \left( EA e_\text{o} + 2w \gamma_{\text{sv}} - \nu_\text{o} \right) \bar{e}_\text{o} \d s  \nonumber \\
&& + \Bigl[ ( \nu_\text{i}-T-\eta) \bar{x}\Bigr]_0^D -\int_0^D \nu_\text{i}' \bar{x} \d s
+ \Bigl[ ( \nu_\text{o}-T) \bar{x}\Bigr]_D^L -\int_D^L \nu_\text{o}' \bar{x} \d s  \nonumber \\
&& + \bar{D} \left[
\frac{1}{2} EA e_\text{i}^2(D) + 2w \gamma_{\text{s} \ell}(1+e_\text{i}(D)) - (T+\eta) x'_\text{i}(D)  \right. \nonumber \\
&& - \left. \frac{1}{2} EA e_\text{o}^2(D) - 2w \gamma_{\text{sv}}(1+e_\text{o}(D)) + T x'_\text{o}(D)
 \right]
\label{equa:1st-varia-big}
\end{eqnarray}
where we have used $\int_0^{D+\varepsilon \bar{D}} f(s) \d s=\int_0^D f(s) \d s+ \varepsilon \bar{D} f(D) + O(\varepsilon^2)$.
Note that since the position $x(s)$, and hence $\bar{x}(s)$, has to be continuous, we do not use any subscript for $x(s)$ or $\bar{x}(s)$.

We first examine the conditions for the first variation (\ref{equa:1st-varia-big}) to vanish for all $\bar{x}(s)$. From the boundary condition $x(0)=0$, we have $\bar{x}(0)=0$, but $\bar{x}(D)$ and $\bar{x}(L)$ are arbitrary. The conditions are then
\begin{subequations}
\label{sys:conditions-xbar}
\begin{eqnarray}
\nu_\text{i}'(s) &=& 0 \mbox{~ and ~} \nu_\text{o}'(s) = 0  \\
\nu_\text{o}(L)&=&T \\
\nu_\text{o}(D) - \nu_\text{i}(D) + \eta &=0 \label{EQ24}
\end{eqnarray}
\end{subequations}
Here again we interpret $\nu_\text{i}(s)$ and $\nu_\text{o}(s)$ as the internal force in the beam, which experiences a jump of amplitude $\eta$ at the contact line.
The conditions for the first variation (\ref{equa:1st-varia-big}) to vanish for all $\bar{\beta}$ and $\bar{r}$ are:
\begin{subequations}
\label{sys:conditions-beta-r}
\begin{eqnarray}
\mu &=& \gamma_{\ell \text{v}}/r \label{EQ21} \\
\eta &=& - 2 w \gamma_{\ell \text{v}} \cos \beta \label{EQ22}
\end{eqnarray}
\end{subequations}
where we see that the Lagrange multiplier $\mu$, enforcing volume, is the Laplace pressure inside the liquid drop $\gamma_{\ell \text{v}}/r$. From (\ref{EQ24}) and (\ref{EQ22}) the Lagrange multiplier $\eta$, associated to the constraint on the position of the contact line, is interpreted as the external force applied on the beam at the contact line (from the liquid and vapor phases). For each drop the force on the beam is of intensity $\gamma_{\ell \text{v}} w$ and is oriented along the liquid-vapor interface. This results is not changed in the case where surface energies depend on strains, see Appendix \ref{appendiceA}.
The conditions for the first variation (\ref{equa:1st-varia-big}) to vanish for all $\bar{e}_\text{i}(s)$ and $\bar{e}_\text{o}(s)$ are:
\begin{subequations}
\label{new-const-rel}
\begin{eqnarray}
\nu_\text{i}(s) &=& EA e_\text{i}(s) + 2w \gamma_{\text{s} \ell} \label{EQ18} \\
 \nu_\text{o}(s) &=& EA e_\text{o}(s) + 2w \gamma_{\text{sv}} \label{EQ17}
\end{eqnarray}
\end{subequations}
which are interpreted as constitutive relations between extension $e_\text{i}$ (resp. $e_\text{o}$) and internal force $\nu_\text{i}$ (resp. $\nu_\text{o}$) in each region of the beam.
Finally the condition for the first variation (\ref{equa:1st-varia-big}) to vanish for all $\bar{D}$ is 
\begin{subequations}
\begin{eqnarray}
\left[ 1+ e_\text{i}(D) \right]^2 - \left[1 + e_\text{o}(D)\right]^2 &=&0 \; \; \mbox{ or }  \label{EQfirst} \\
\left[ e_\text{i}(D) - e_\text{o}(D)\right] \, \left[ e_\text{i}(D) + e_\text{o}(D) +2 \right] &=& 0 \label{EQsecond}
\end{eqnarray}
\end{subequations}
As the vanishing of the second term of (\ref{EQsecond}) would require oversized extension values, this equality can only be fulfilled if
\begin{equation}
e_\text{i}(D) - e_\text{o}(D) = 0 \label{EQ19}
\end{equation}
that is the extension is continuous at the contact line.
Considering (\ref{EQ24}), (\ref{EQ22}), (\ref{EQ18}), (\ref{EQ17}), and (\ref{EQ19}), we obtain
\begin{equation}
\gamma_{\text{s} \ell} - \gamma_{\text{sv}} + \gamma_{\ell \text{v}} \cos \beta = 0 \label{YD}
\end{equation}
which means that the wetting angle satisfies Young-Dupr\'e relation.

We conclude that in the case where an extensible beam is in capillary interaction with a liquid drop we have the three following properties: $(i)$ the constitutive relation linking internal force and extension is modified and includes a surface tension term, $(ii)$ the internal force inside the beam experiences a jump at the contact line, corresponding to the force coming from the liquid-vapor interface, this force being oriented along the interface, and $(iii)$ the extension of the beam is continuous at the triple line.

\section{Beams experiencing both bending and extension} \label{section:flexion-ext}
\begin{figure}[ht]
    \centering
    \includegraphics{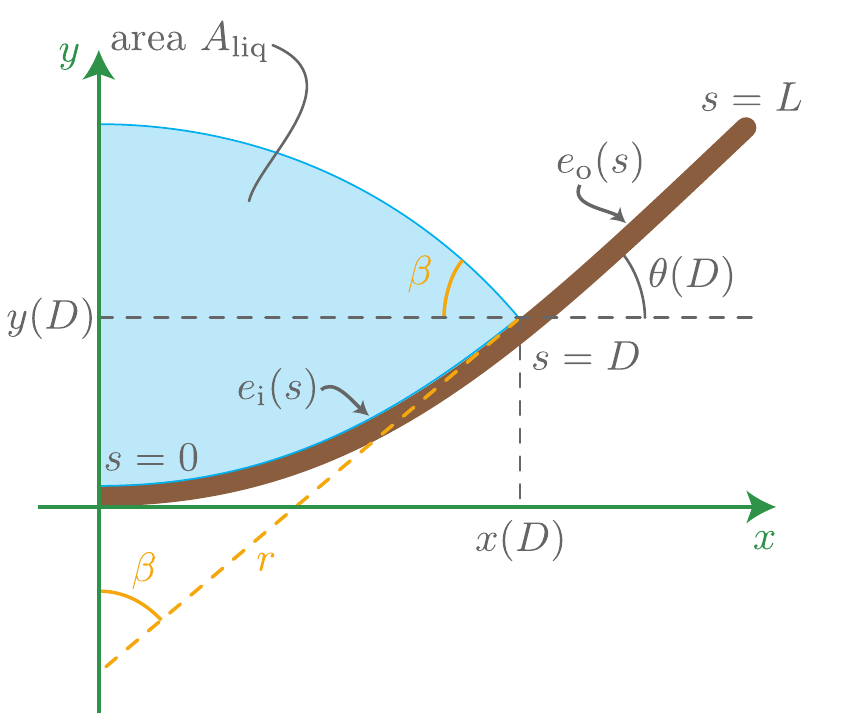}
    \caption{A beam in interaction with a liquid drop. In the deformed state both bending and extension are present.}
    \label{fig-flex-ext}
\end{figure}

%
We now consider the case where a liquid drop sits on the top of a flexible and extensible beam and we look for equilibrium equations, see Figure \ref{fig-flex-ext}. 
As in the former section, we use a beam of rectangular cross-section (thickness $h$, width $w$) and set $P=2w$ and $A=hw$. We focus on one half of the system, $s \in [0;L]$, and work with the following boundary conditions:
\begin{equation}
x(0)=0 \, , \quad y(0)=0 \, , \quad \theta(0)=0 \label{equa:BC-s0}
\end{equation}
As in the previous sections, $s$ is the arc-length of the beam in its reference state. Consequently once the beam is deformed, it may no longer have total contour-length $L$.
To the stretching energy $V_\text{e}$, Eq. (\ref{equa-stretch-energy}), we add the bending energy:
\begin{equation}
V_\kappa= \frac{1}{2} EI \int_{0}^{D} \left[ \theta_\text{i}'(s) \right]^2 ds + \frac{1}{2} EI \int_{D}^{L} \left[ \theta_\text{o}'(s) \right]^2 ds
\end{equation}
where $EI$ is the bending rigidity of the strip ($E$ is Young's modulus of the beam material, and $I=h^3 w / 12$ is the second moment of area of the section of the beam).
Ignoring constant terms, the sum of the interfaces energies is \footnote{Although it is not a constant term, adding the surface energy of the lower side of the beam does not change the present results.}:
\begin{eqnarray}
V_\text{s}= w \gamma_{\text{s} \ell} \int_0^D  [1+e_\text{i}(s)] \d s +  w \gamma_{\text{sv}} \int_D^L [1+e_\text{o}(s)] \d s +  w \gamma_{\ell \text{v}} r \beta
\end{eqnarray}
We minimize $V_\text{e}+V_\kappa+V_\text{s}$ under the following constraints. First the liquid volume ${\cal V}= A_{\mathrm{liq}} w$ being fixed we have:
\begin{equation}
A_{\mathrm{liq}} =\frac{1}{2}  r^2 \left(\beta -  \sin \beta \cos  \beta \right) + x(D) \, y(D) - \int_{x(0)}^{x(D)} y \, dx \label{equa:fixed_volume}
\end{equation}
Second, we still have the geometric constraint (\ref{equa-xd-contraint}).
And finally the relations linking the unknown functions $x(s),y(s),\theta(s), e_\text{i}(s),e_\text{o}(s)$ are for $s$ in $(0,D)$: $x_\text{i}'(s)=(1+ e_\text{i}(s)) \cos \theta(s)$ and $y_\text{i}'(s)=(1+ e_\text{i}(s)) \sin \theta(s)$, and for $s$ in $(D,L)$: $x_\text{o}'(s)=(1+ e_\text{o}(s)) \cos \theta(s)$ and $y_o'(s)=(1+ e_\text{o}(s)) \sin \theta(s)$.

We therefore compute the first variation of:
\begin{eqnarray}
{\cal L}&=& V_\kappa+V_\text{e}+V_\text{s} - \eta  \left[ \int_0^D x_\text{i}' \d s - r \sin \beta \right] \nonumber \\  
&& - \mu w \left[ \frac{r^2}{2} \left(\beta - \sin \beta \cos \beta \right) + \int_0^D x_\text{i}' \d s \, \times \,  \int_0^D y_\text{i}' \d s - \int_{0}^{D} y \, x_\text{i}' \, \d s \right]  \nonumber \\
&&  
+ \int_{0}^{D} \nu_\text{i}(s) \,  \left[x_\text{i}'-(1+ e_\text{i}) \cos \theta \right] \d s 
+ \int_{D}^{L} \nu_\text{o}(s) \,  \left[x_\text{o}'-(1+ e_\text{o}) \cos \theta \right] \d s \nonumber \\
&& 
+ \int_{0}^{D} \lambda_\text{i}(s) \,  \left[y_\text{i}'-(1+ e_\text{i}) \sin \theta \right] \d s
+ \int_{D}^{L} \lambda_\text{o}(s) \,  \left[y_o'-(1+ e_\text{o}) \sin \theta \right] \d s 
\label{equa:lagrangian2}
\end{eqnarray}
where ${\cal L}={\cal L}(x,y,e_\text{i},e_\text{o},\theta,\beta,r,D)$.
Computing the first derivative, we obtain:
\begin{eqnarray}
\left. \frac{\d {\cal L}}{\d \varepsilon} \right|_{\varepsilon=0}&=&
EI \int_{0}^{D} \theta_\text{i}' \, \bar{\theta}_\text{i}' \d s + \frac{1}{2} EI \bar{D} \left[ \theta_\text{i}'(D) \right]^2 +
EI \int_{D}^{L} \theta_\text{o}' \, \bar{\theta}_\text{o}' \d s - \frac{1}{2} EI \bar{D} \left[ \theta_\text{o}'(D) \right]^2 \nonumber \\
&&
+EA \int_{0}^{D} e_\text{i} \, \bar{e}_\text{i} \d s + \frac{1}{2} EA \bar{D} e_\text{i}^2(D)+
EA \int_{D}^{L} e_\text{o} \, \bar{e}_\text{o} \d s - \frac{1}{2} EA \bar{D}  e_\text{o}^2(D)  \nonumber \\
&&
+ w \gamma_{\text{s} \ell} \int_0^D  \bar{e}_\text{i} \d s 
+ w \gamma_{\text{s} \ell} \bar{D} [1+e_\text{i}(D)] 
+ w \gamma_{\text{sv}} \int_D^L \bar{e}_\text{o} \d s 
-  w \gamma_{\text{sv}} \bar{D} [1+e_\text{o}(D)] 
 \nonumber \\
&&
+  w \gamma_{\ell \text{v}} (\bar{r} \beta+r \bar{\beta}) - \mu w \left[ r \bar{r} \left(\beta - \frac{1}{2} \sin 2 \beta \right) 
+ \frac{\bar{\beta} r^2}{2}  \left(1 - \cos 2 \beta \right) \right] \nonumber \\ 
&&- \mu w \left[ \int_0^D \bar{x}_\text{i}' \d s \, \times \,  \int_0^D y_\text{i}' \d s  + \int_0^D x_\text{i}' \d s \, \times \,  \int_0^D \bar{y}_\text{i}' \d s + \bar{D} x_\text{i}'(D) y(D) + \bar{D} x(D) y_\text{i}'(D)  \right]  \nonumber \\
&&+ \mu w \left[  \int_{0}^{D} \bar{y} \, x_\text{i}' \, \d s +  \int_{0}^{D} y \, \bar{x}_\text{i}' \, \d s 
+ \bar{D} y(D) x_\text{i}'(D) \right]  \nonumber \\
&&
- \eta  \left[ \int_0^D \bar{x}_\text{i}' \d s +\bar{D} x_\text{i}'(D) - \bar{r} \sin \beta - r \bar{\beta} \cos \beta \right] \nonumber \\
&&  
+ \int_{0}^{D} \nu_\text{i}(s) \,  \left[\bar{x}_\text{i}' - \bar{e}_\text{i} \cos \theta + \bar{\theta}(1+e_\text{i})\sin \theta \right] \d s \nonumber \\
&&
+ \int_{D}^{L} \nu_\text{o}(s) \,  \left[\bar{x}_\text{o}' - \bar{e}_\text{o} \cos \theta + \bar{\theta}(1+e_\text{o})\sin \theta \right] \d s \nonumber \\
&& 
+ \int_{0}^{D} \lambda_\text{i}(s) \,  \left[\bar{y}_\text{i}' - \bar{e}_\text{i} \sin \theta - \bar{\theta}(1+e_\text{i})\cos \theta \right] \d s \nonumber \\
&&
+ \int_{D}^{L} \lambda_\text{o}(s) \,  \left[\bar{y}_\text{o}' - \bar{e}_\text{o} \sin \theta - \bar{\theta}(1+e_\text{o})\cos \theta \right] \d s
\label{equa:1st-variation-expression}
\end{eqnarray}
As $x(s)$, $y(s)$, and $\theta(s)$ are assumed to be continuous so are their variations $\bar{x}(s)$, $\bar{y}(s)$, and $\bar{\theta}(s)$. Consequently these variables do not carry any subscript. 
As in the previous section, the conditions for the first variation (\ref{equa:1st-variation-expression}) to vanish for all $\bar{\beta}$ and $\bar{r}$ yields:
\begin{equation}
\mu = \gamma_{\ell \text{v}}/r \: \mbox{ and } \: \eta = - w \gamma_{\ell \text{v}} \cos \beta \label{EQneweta}
\end{equation}
Requiring (\ref{equa:1st-variation-expression}) to vanish for all $\bar{x}$ yields, after integration by parts:
\begin{equation}
\Bigl[ \bigl( - \mu w \, \left(y(D)-y \right) - \eta  + \nu_\text{i} \bigr) \, \bar{x} \, \Bigr]_0^D  
- \int_0^D \left( \mu w y_\text{i}' + \nu_\text{i}' \right) \bar{x} \, ds 
+ \Bigl[ \nu_\text{o} \, \bar{x} \, \Bigr]_D^L -\int_D^L  \nu_\text{o}' \, \bar{x} \, ds =0
\end{equation}
The fact that we have $\bar{x}(0)=0$, but arbitrary $\bar{x}(D)$ and $\bar{x}(L)$ implies:
\begin{equation}
\nu_\text{o}(L) = 0 \, , \; \;  \nu_\text{o}(D) - \nu_\text{i}(D) = - \eta  \, , \; \;
\nu'_\text{o}(s) = 0 \, , \; \;  \nu'_\text{o} (s)= - \mu w y_\text{i}'(s) \label{equa:equilibre_force_x}
\end{equation}
Requiring (\ref{equa:1st-variation-expression}) to vanish for all $\bar{y}$ similarly implies:
\begin{equation}
\lambda_\text{o}(L) = 0 \, , \; \;  \lambda_\text{o}(D) - \lambda_\text{i}(D) = - \mu w x(D) \, , \; \; \lambda'_o(s) = 0 \, , \; \;  \lambda'_i(s) =  \mu w x_\text{i}'(s) \label{equa:equilibre_force_y}
\end{equation}
%
Here again we identify $\bm{N}_i(s)=(\nu_\text{i}(s),\lambda_\text{i}(s))$ and $\bm{N}_o(s)=(\nu_\text{o}(s),\lambda_\text{o}(s))$ as the internal force in the beam.
We see that Laplace pressure $\mu$ generates an outward normal distributed force $\mu w (y_\text{i}',-x_\text{i}')$ that causes the internal force $\bm{N}_i(s)$ to vary.
In addition, using (\ref{EQneweta}) we see that at the contact line $s=D$ the force experiences a jump, of amplitude $w\gamma_{\ell \text{v}}$ and oriented along the liquid-air interface:
\begin{equation}
\bm{N}_o(D)-\bm{N}_i(D) = - \gamma_{\ell \text{v}} w \, 
\begin{pmatrix} -\cos \beta \\ \sin \beta \end{pmatrix}
\end{equation}
Requiring (\ref{equa:1st-variation-expression}) to vanish for all $\bar{\theta}$ yields, after integration by parts
\begin{eqnarray}
EI  \left[ \theta_\text{i}' \, \bar{\theta} \, \right]_0^D 
&+&\int_0^D \left[ - EI \theta_\text{i}'' + (1+e_\text{i}) (\nu_\text{i} \sin \theta - \lambda_\text{i} \cos \theta) \right] \bar{\theta} \d s  \nonumber\\
&+& EI \left[ \theta'_\text{o} \, \bar{\theta} \, \right]_D^L
+\int_D^L \left[ - EI \theta_\text{o}'' + (1+e_\text{o}) (\nu_\text{o} \sin \theta - \lambda_\text{o} \cos \theta) \right] \bar{\theta} \d s
=0    \label{eq:theta}
\end{eqnarray}
Boundary condition $\theta(0)=0$  imposes $\bar{\theta}(0)=0$, but $\bar{\theta}(D)$ and $\bar{\theta}(L)$ are arbitrary. Consequently (\ref{eq:theta}) yields
\begin{subequations}
\begin{eqnarray}
\theta_\text{i}'(D)&=&\theta_\text{o}'(D) \\
\theta_\text{o}'(L)&=&0 \\ 
EI\theta_\text{i}''(s) &=& (1+e_\text{i}) \left[ \nu_\text{i} \sin \theta - \lambda_\text{i} \cos \theta \right] \\
EI\theta_\text{o}''(s) &=& (1+e_\text{o}) \left[ \nu_\text{o} \sin \theta - \lambda_\text{o} \cos \theta \right] 
\end{eqnarray}
\end{subequations}
That is the curvature $\theta'(s)$ (and hence the bending moment) is continuous as $s$ goes through $D$ and it vanishes at the $s=L$ extremity. Moreover we recognize in the last two equations the moment equilibrium equations along the beam.
Requiring (\ref{equa:1st-variation-expression}) to vanish for all $\bar{e}_\text{i}(s)$ and $\bar{e}_\text{o}(s)$ yields the two following relations
\begin{subequations}
\begin{eqnarray}
EA \, e_\text{i}(s) + w \gamma_{\text{s} \ell} &=& \nu_\text{i} \cos \theta + \lambda_\text{i} \sin \theta \\
EA \, e_\text{o}(s) + w \gamma_{\text{sv}} &=& \nu_\text{o} \cos \theta + \lambda_\text{o} \sin \theta
\end{eqnarray}
\end{subequations}
that we interpret as constitutive relations linking the extension $e(s)$ to the internal tension $\bm{N}(s) \cdot \bm{t}(s)$ where $\bm{t}(s)$ is the tangent to the beam $\bm{t}(s)=(\cos \theta(s) , \sin \theta(s))$.
Using (\ref{equa:equilibre_force_x}) and (\ref{equa:equilibre_force_y}) we see that $EA e_\text{o}(s) = - w \gamma_{\text{sv}}$ for all $s$, and that
\begin{equation}
\frac{EA}{w}\, \left[e_\text{o}(D)-e_\text{i}(D) \right] =  \gamma_{\text{s} \ell}-\gamma_{\text{sv}} + \gamma_{\ell \text{v}} 
\cos \left[ \theta(D)+\beta \right]
\label{EQ46}
\end{equation}
Finally, requiring (\ref{equa:1st-variation-expression}) to vanish for all $\bar{D}$ yields
\begin{equation}
\frac{1}{2} EA e_\text{i}^2 - \frac{1}{2} EA e_\text{o}^2 + w \gamma_{\text{s} \ell} (1+e_\text{i}(D)) -w \gamma_{\text{sv}} (1+e_\text{o}(D)) - \mu w x(D) y_\text{i}'(D) - \eta x_\text{i}'(D) =0
\end{equation}
Using (\ref{equa-xd-contraint}), (\ref{EQneweta}), and (\ref{EQ46}) this is simplified to
\begin{subequations}
\begin{eqnarray}
\left[ 1+ e_\text{i}(D) \right]^2 - \left[1 + e_\text{o}(D)\right]^2 &=&0 \; \; \mbox{ or } \\
\left[ e_\text{i}(D) - e_\text{o}(D)\right] \, \left[ e_\text{i}(D) + e_\text{o}(D) +2 \right] &=& 0 \label{EQsecondbis}
\end{eqnarray}
\end{subequations}
Consequently we see that even in the presence of bending, the extension is continuous as $s$ go through $D$ and the young-Dupr\'e relation holds for the wetting angle:
\begin{eqnarray}
e_\text{i}(D) - e_\text{o}(D) &=& 0  \\
\gamma_{\text{s} \ell} - \gamma_{\text{sv}} + \gamma_{\ell \text{v}} \cos \left[ \theta(D)+\beta \right] &=& 0
\end{eqnarray}

%
%
%
\section{Conclusion} \label{section:conclusion}
%
%
%
%
We have shown that the constitutive relation between tension and extension in an elastic beam is altered by surface energies, leading to a modification of Hooke's law.
In a setup where three phases are involved (solid, liquid, and vapor) we have shown the following properties, in the case where surface energies do not depend on the strain state in the solid, $(p_1)$ the extension in the beam is continuous at the triple line, $(p_2)$ the wetting angle satisfies Young-Dupr\'e relation, and $(p_3)$ the external force applied on the beam is along the liquid-vapor interface. This last property $(p_3)$ has been shown to hold even if strain dependence is introduced in the surface energies.
Properties $(p_2)$ and $(p_3)$ have already been established in \cite{Neukirch-Antkowiak:The-bending-of-an-elastic-beam:2013} in the pure bending case, and we have verified here that in a setup where bending and extension are both present, these properties still hold.
The new constitutive relations (Eq.(\ref{new-const-rel}), or Eq.(\ref{new-new-const-rel})) between tension and extension compel one to be careful when analyzing experimental results, as a jump in extension no longer implies the same jump in tension.
For example in Section \ref{section:force-jump} we find continuous extension and a discontinuous tension.
Experimental results reported in \cite{Marchand-Das:Capillary-Pressure-and-Contact:2012} indicate the extension changes sign at the contact line. The present derivations show that this is not possible under the classical hypotheses used in Section \ref{section:force-jump}, and that for example strain dependence of surface energies have to be invoked, as also stated in \cite{Weijs-Andreotti:Elasto-capillarity-at-the-nanoscale:-on-the-coupling:2013}. Moreover in order for $e_\text{i}(D)$ and $e_\text{o}(D)$ to have opposite sign, we see that the derivative $\gamma_{\text{s} \ell}'(e)$ and $\gamma_{\text{sv}}'(e)$ have to be of the same order as $\gamma_{\text{s} \ell}$ and $\gamma_{\text{sv}}$. 
The sign change of the extension at the contact line was attributed to the presence of a tangential component in the force at the contact line \cite{Marchand-Das:Capillary-Pressure-and-Contact:2012},  derived from a microscopic model of capillarity in \cite{Das-Marchand:Elastic-deformation-due-to-tangential:2011}, and observed in Molecular Dynamics simulations \cite{Seveno2013Youngs-Equation,Weijs-Andreotti:Elasto-capillarity-at-the-nanoscale:-on-the-coupling:2013}.
Nevertheless present results indicate that no tangential component, in the external force applied on the beam at the contact line, is needed in order to have a sign change of the extension.

As a matter of fact this external force is here found to classically lie along the liquid-vapor interface even in the case of strain-dependent surface energies.
\begin{acknowledgments}
S.N. thanks Bruno Andreotti for discussions. We are grateful to R\'egis Wunenburger for his remarks on surface stresses.
The present work was supported by ANR grant  ANR-09-JCJC-0022-01.
Financial support from `La Ville de Paris - Programme \'Emergence' is also  acknowledged. \end{acknowledgments}

\appendix
\section{Strain-dependent surface energy} \label{appendiceA}
\begin{figure}[ht]
    \centering
    \includegraphics{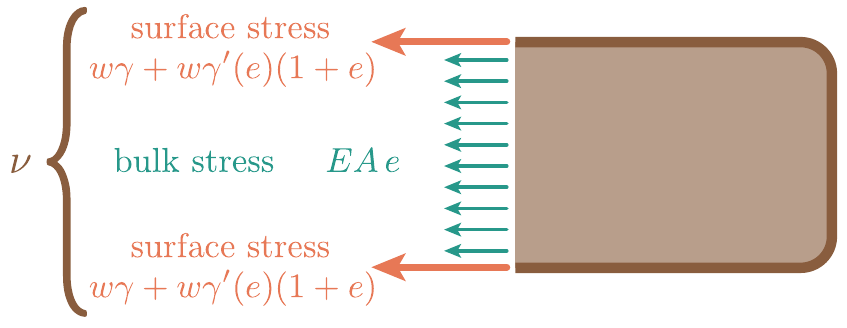}
    \caption{Decomposition of the beam internal tension into bulk and surface stresses.}
    \label{fig-shu}
\end{figure}

We investigate here how the results of Section \ref{section:force-jump} change if surface energies $\gamma_{\text{s} \ell}$ and $\gamma_{\text{sv}}$ depend on the strain of the elastic material, that is $\gamma_{\text{s} \ell}=\gamma_{\text{s} \ell}(e_\text{i})$ and $\gamma_{\text{sv}}=\gamma_{\text{sv}}(e_\text{o})$ \cite{Muller-Saul:Elastic-effects-on-surface:2004,Weijs-Andreotti:Elasto-capillarity-at-the-nanoscale:-on-the-coupling:2013}.
First, results in (\ref{sys:conditions-xbar}) and (\ref{sys:conditions-beta-r}) remain unchanged. Consequently the Lagrange multiplier $\nu$ is still interpreted as the internal force, and the force applied on the beam at the contact line is still of intensity $\gamma_{\ell \text{v}} w$ and still oriented along the liquid-vapor interface, in contradiction to what is argued in \cite{Weijs-Andreotti:Elasto-capillarity-at-the-nanoscale:-on-the-coupling:2013}.
Second, the constitutive relation between tension and extension, formerly (\ref{EQ18}) and (\ref{EQ17}), now comprises one more term:
\begin{subequations}
\label{new-new-const-rel}
\begin{eqnarray}
\nu_\text{i}(s) &=& EA e_\text{i}(s) + 2w \gamma_{\text{s} \ell} + 2w \gamma_{\text{s} \ell}'(e_\text{i}) [1+e_\text{i}(s)]\label{EQA2} \\
 \nu_\text{o}(s) &=& EA e_\text{o}(s) + 2w \gamma_{\text{sv}} + 2w \gamma_{\text{sv}}'(e_\text{o}) [1+e_\text{o}(s)]\label{EQA3}
\end{eqnarray}
\end{subequations}
As depicted in Fig.~\ref{fig-shu} these relations are the beam mechanics version of Shuttleworth's equation \cite{Shuttleworth:The-Surface-Tension-of-Solids:1950}. 
In this case equation (\ref{EQfirst}) is modified and now reads
\begin{equation}
\left[ 1+ e_\text{o}(D) \right]^2  \left( 1+  \frac{4\gamma_{\text{sv}}'(e_\text{o}(D))}{EA/w} \right)=
 \left[1 + e_\text{i}(D)\right]^2 \left( 1+  \frac{4\gamma_{\text{s} \ell}'(e_\text{i}(D))}{EA/w} \right)
\end{equation}
and shows that the extension now has a discontinuity at the contact line.
Third, relation (\ref{YD}), giving the wetting angle, is also changed and becomes
\begin{equation}
\gamma_{\text{s} \ell} - \gamma_{\text{sv}} + \gamma_{\ell \text{v}} \cos \beta = 
\frac{EA}{2w} \left(e_\text{o}(D)-e_\text{i}(D) \right) +
\gamma_{\text{sv}}'(e_\text{o}) [1+e_\text{o}(D)] -\gamma_{\text{s} \ell}'(e_\text{i}) [1+e_\text{i}(D)]
\end{equation}
If we now assume that $(EA/w) \, e $  and $\gamma'(e)$  are both of order $\gamma$, and if we neglect $O(e^2)$ terms we find:
\begin{eqnarray}
e_\text{o}(D)-e_\text{i}(D) &=& \frac{2}{EA/w} \, \left[ \gamma_{\text{s} \ell}'(e_\text{i}) - \gamma_{\text{sv}}'(e_\text{o}) \right] + O(e^2) \\
\gamma_{\text{s} \ell} - \gamma_{\text{sv}} + \gamma_{\ell \text{v}} \cos \beta &=& O(e^2)
\end{eqnarray}
as also found in \cite{Weijs-Andreotti:Elasto-capillarity-at-the-nanoscale:-on-the-coupling:2013}.

\section{Classical beam equations} \label{appendixB}
\begin{figure}[ht]
    \centering
    \includegraphics{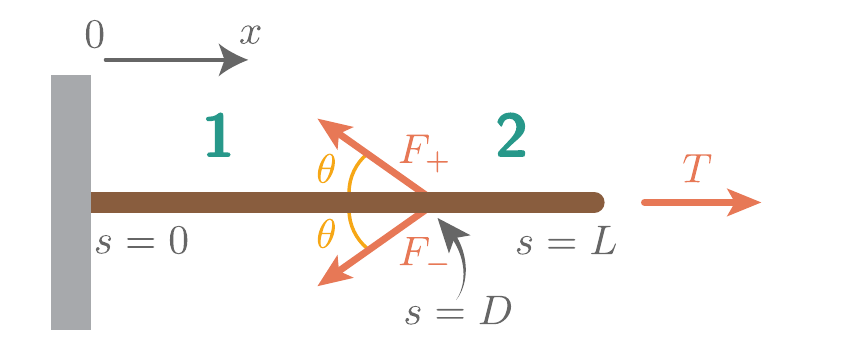}
    \caption{A beam subjected to an external tension $T$ and two compressive forces $\bm{F}_\pm$.}
    \label{fig2appendix}
\end{figure}
In order to illustrate our interpretation of the Lagrange multipliers $\nu$ and $\eta$ in Section \ref{section:ext-seule}, we recall here the classical equilibrium equations for a  beam  subjected to external forces. In Fig.~\ref{fig2appendix} we show a beam anchored at $s=0$ and subjected to an external tension $T$ at $s=L$. Moreover two external forces are applied at $s=D$:  $\bm{F}_\pm=F \, (-\cos \theta , \pm \sin \theta)$. In this case equations for the longitudinal internal beam force $N(s)$ are:
\begin{subequations}
\label{sys:equil-force-beam}
\begin{eqnarray}
N_1'(s) &=& 0 \mbox{~ and ~} N_2'(s) = 0  \label{equa-force-eq} \\
N_2(L)&=&T \label{equa-force-bc} \\
N_2(D) &-&N_1(D) - 2 F \cos \theta =0 \label{equa-force-jump}
\end{eqnarray}
\end{subequations}
where the index 1 (respectively 2) refers to the beam region $s \in (0;D)$ (resp. $s \in (D;L)$). Eq.~(\ref{equa-force-eq}) is the local force equilibrium for a beam with no distributed load (e.g. gravity). Eq.~(\ref{equa-force-bc}) is the boundary condition stating that the external applied force at $s=L$ is tension $T$. Eq.~(\ref{equa-force-jump}) is the force equilibrium at $s=D$. Comparison of equations~(\ref{sys:equil-force-beam}) with equations~(\ref{sys:conditions-xbar}) and (\ref{EQ22}) naturally leads to the interpretation of the Lagrange multiplier $\nu$  in Section  \ref{section:force-jump} as the internal force in the beam, of $w \, \gamma_{\ell \text{v}}$ as the intensity of the external force at $s=D$ and of $\beta$ as its orientation.

\bibliography{soft-beams}
\end{document}